%% file: emse20-sampling.tex
\RequirePackage{fix-cm}
\documentclass[smallextended]{svjour3}       
\smartqed  
\usepackage{graphicx}
\journalname{Empirical Software Engineering}

\usepackage{booktabs} 
\usepackage{graphicx}
\usepackage{tabularx}
\usepackage{multirow}
\usepackage{enumitem}
\usepackage{calc}
\usepackage{tabularx}

\usepackage{hyperref}

\usepackage{natbib}
\setcitestyle{authoryear}

\usepackage{todonotes}

\begin{document}

\title{Sampling in Software Engineering Research: A~Critical~Review~and~Guidelines}


\author{Sebastian Baltes \and Paul Ralph}


\institute{S. Baltes \at
              University~of~Adelaide, Adelaide, Australia\\
              \email{sebastian.baltes@adelaide.edu.au}
           \and
           P. Ralph \at
              Dalhousie~University, Halifax, Canada\\
              \email{paulralph@dal.ca}
}

\date{Received: date / Accepted: date}

\maketitle

\begin{abstract}
Representative sampling appears rare in empirical software engineering research. Not all studies need representative samples, but a general lack of representative sampling undermines a scientific field. This article therefore reports a critical review of the state of sampling in recent, high-quality software engineering research. The key findings are: (1) random sampling is rare; (2) sophisticated sampling strategies are very rare; (3) sampling, representativeness and randomness often appear misunderstood. These findings suggest that \textit{software engineering research has a generalizability crisis}. To address these problems, this paper synthesizes existing knowledge of sampling into a succinct primer and proposes extensive guidelines for improving the conduct, presentation and evaluation of sampling in software engineering research. It is further recommended that while researchers should strive for more representative samples, disparaging non-probability sampling is generally capricious and particularly misguided for predominately qualitative research.
\end{abstract}

\input{tex/1_introduction}
\input{tex/2_background}

\input{tex/3_method}
\input{tex/4_results}
\input{tex/5_recommendations}
\input{tex/7_relatedWork}
\input{tex/6_conclusion}

\begin{acknowledgements}
The authors would like to thank Klaas Stol for encouraging us to write this paper.
This project was supported by the National Sciences and Engineering Research Council of Canada (Grant No. RGPIN-2020-05001)
\end{acknowledgements}

%
%

\bibliographystyle{spbasic}      
\bibliography{bib}

\end{document}

%% file: tex/1_introduction.tex
\section{Introduction}
\label{introduction}

Most research involves selecting some of many possible items to study; that is, \textit{sampling}. Sampling is crucial not only for positivist research (e.g. questionnaires, experiments, positivist case studies) but also for interpretivist and constructivist research (e.g. interview studies, grounded theory, interpretivist case studies). 


In present-day epistemology, positivism refers to various ideological and philosophical positions usually associated with quantitative research, such as belief in objective reality and attempting to support a priori theories with  (usually quantitative) empirical evidence. Sampling is crucial for positivist studies because it determines external validity.

Meanwhile, social constructivism is the idea that much of our reality is socially or culturally constructed. This leads to interpretivism: the view that the methods of natural science are insufficient or inappropriate for understanding social reality \textit{because} it is socially and culturally constructed. Interpretivists believe that each situation is unique and concepts from one context do not necessarily generalize to others.  Both schools are associated with building (more so than testing) context-dependent, middle-range theories. The goal of interpretivists and constructivists is not statistically generalizing from samples to populations---they sometimes  do not even use the term ``sampling''. However, sampling is still crucial for them because selecting poor research sites hinders data collection and focusing on the wrong topics undermines theory building.

We were motivated to write this paper by three observations concerning the difficulty of sampling for software engineering (SE) research:

\begin{enumerate}
    \item For many SE phenomena, there is no suitable \emph{sampling frame}; that is, a list from which to draw a sample.
    \item Some SE studies adopt poorly understood sampling strategies such as random sampling from a \emph{non-representative surrogate population}.
    \item Many SE articles evince deep misunderstandings of \textit{representativeness}---the key criteria for assessing sampling in positivist research (see Section~\ref{represent}).
\end{enumerate}


The purpose of this paper is therefore as follows.

\smallskip
    {\narrower \noindent \textit{\textbf{Purpose:} (1) Providing a detailed, SE-specific primer on sampling concerns and techniques; (2) investigating the state of sampling in SE research; (3) providing guidelines for improving sampling in SE research.}\par}
\smallskip

Please note: This paper discusses common problems and errors in existing research (e.g. misusing the term \textit{random}, overstating sample representativeness). However, we intentionally do not cite specific examples of papers that make these errors because our goal is to foster consensus not division. We do not wish to cast aspersions on individual researchers for endemic problems. Anyone who reads widely in SE research should recognize the ubiquity of the problems described below. 


%% file: tex/2_background.tex
\section{Sampling: A Primer} 
\label{background}

For our purposes, \textit{sampling} is the process of selecting a group of items to study (a \textit{sample}) from a (usually larger) group of items of interest. The group of items of interest is called the \textit{population}, and the (usually incomplete) list of items in the population is called the \textit{sampling frame}. The degree to which a sample's properties (of interest) resemble those of the population is the sample's \textit{representativeness} (see Section \ref{represent} for more). 

Researchers have developed diverse approaches for selecting samples from populations. There is no single best technique or hierarchy of better and worse techniques. Different sampling approaches are appropriate for different purposes in different circumstances. While there are dozens of sampling strategies, this section focuses on those most applicable to SE research. 


The following exposition on sampling is primarily synthesized from several textbooks \citep{cochran2007sampling,henry1990practical,trochim2001research} and extends similar primers in the SE literature \citep[e.g.][]{de2015investigating,kitchenham2008personal}. A good grasp of the sampling strategies and issues dissected in this section is necessary to understand the study that follows. Table \ref{tab_primerSummary} summarizes the sampling approaches.

\input{tables/primerSummary}

\subsection{Non-probability sampling}
Non-probability sampling includes all of the sampling techniques that do not employ randomness. This section discusses four kinds of non-probability sampling: convenience, purposive, referral-chain (or snowball) sampling, and respondent-driven sampling.

While some (positivist) methodologists \citep[e.g.][]{van2015aiming} present non-probability sampling as intrinsically inferior to probability sampling, non-probability sampling is preferable for some purposes in some contexts (see Section \ref{represent}).  

\subsubsection{Convenience sampling} 

In convenience sampling, items are selected based on \textit{availability or expedience}. When we select people or artifacts to study arbitrarily, or based on them being nearby, available or otherwise easy to study, we adopt convenience sampling. Convenience sampling is controversial because it is very popular despite threatening generalizability  \cite[cf.][]{arnett2008neglected,henrich2010weirdest}. The key advantages of convenience sampling are: (1) speed, (2) low cost and (3) no need for a sampling frame. 

\subsubsection{Purposive sampling} 

In purposive sampling, items are selected according to some logic or strategy, carefully but not randomly  \citep{patton2014qualitative}. Guidelines for purposive sampling are often provided in the context of selecting sites or data sources for predominately qualitative, interpretivist research \cite[e.g.][]{miles2014qualitative}. While non-probability samples can be representative of broader populations, the goal of non-probability sampling is often to find accessible, information-rich cases, sites, organizations or contexts from which researchers can learn about their topic of study \citep{patton2014qualitative}.   

Purposive sampling includes several approaches, for example: 
\begin{enumerate}
\item Studying projects hosted on GitHub\footnote{\url{https://github.com}} because it is popular and has good tool support.
\item Recruiting a panel of experts on a particular topic for a focus group (``expert sampling'').
\item Selecting projects that are as diverse as possible (``heterogeneity sampling'')\footnote{Diversity can be defined along many different axes, gender being one of them~\citep{vasilescu_gender_2015}.}.
\end{enumerate}

The key advantages of purposive sampling are: (1) the researcher can exercise expert judgment; (2) the researcher can ensure representativeness on a specific dimension (see Section \ref{represent}); (3) no sampling frame is needed (e.g. you do not need a list of every software company in your country to select ABC corporation as the site for a case study). The main challenge with purposive sampling is that it is intrinsically subjective and opportunistic.

\subsubsection{Referral-chain (snowball) sampling} 
In referral-chain sampling (also called snowball sampling) sampling, items are selected based on their \textit{relationship to previously selected items}. Snowball sampling is useful when there is no good sampling frame for the population of interest \citep{faugier1997sampling}. For example, there is no comprehensive list of black-hat hackers or software developers who have experienced sexual harassment. However, members of such ``hidden populations'' often know each other. Snowball sampling with human participants therefore works by finding a few individuals in the population, studying them, and having them identify other members of the population. 

In SE, snowball sampling is commonly used in literature reviews to supplement keyword searches. When we begin with an article A, searching the papers A cites is sometimes called \textit{backward snowballing} while searching the papers that cite A is sometimes called \textit{forward snowballing}. We can study software libraries, methods and services (as in service-oriented architecture), in much the same way.

The advantage of snowball sampling is that it helps us to identify items that are not in our sampling frame. However, snowball sampling biases results toward more connected people (or artifacts) and can lead to sampling a small, highly-interconnected subset of a larger population. 


\subsubsection{Respondent-driven sampling\label{rds}} 

Respondent-driven sampling is an advanced form of referral-chain sampling designed to mitigate sampling bias \citep{Heckathorn:1997vw}. It works as follows. 
\begin{enumerate}
\item Begin with diverse initial participants (seeds) who (i) have large social networks, (ii) represent different sub-populations, (iii) do not know each other, and (iv) can influence peers to participate.
\item Have participants recruit, rather than identify, peers. This reduces selection bias by the researcher. 
\item Limit recruitment such that each participant can only recruit a small number of peers (typically three) to prevent highly-connected participants from biasing the sample.
\item Require many (e.g. 20) recruitment waves. This generates longer referral chains, decreasing the risk of oversampling from a highly-connected subset of the population.
\item Prevent individuals from participating more than once.
\item Continue recruitment until the sample reaches \textit{equilibrium}, the point where the distribution of variables of interest is stable.
\item Apply a mathematical model to account for sampling bias \citep[cf.][]{Johnston:2010ko, Heckathorn:1997vw}.
\end{enumerate}

While the details of the mathematical model used are beyond the scope of this paper, more information and tools are available.\footnote{\url{http://respondentdrivensampling.org/}}
Respondent-driven sampling has, however, been criticized for producing optimistic confidence intervals, and generally being ``substantially less accurate than generally acknowledged'' \citep{Goel:2010kr}.

\subsection{Probability sampling}
Probability sampling includes all of the sampling techniques that employ randomness. Here, \textit{random} means that each item in the population has an equal probability of selection \citep{daniel2011sampling}, not selecting items arbitrary or without logic. Standing on a street corner interviewing ``random'' pedestrians, recruiting participants using email or advertising on social networks or assigning participants to experimental conditions in the order in which they arrive at a laboratory are not random. Practically speaking, any selection without using a random number generator is probably not random.\footnote{True random number generation is available from numerous sources, including \url{https://www.random.org/}}

Research on and guidelines for probability sampling are often written from the perspective of positivist questionnaire studies \citep[e.g.][]{kitchenham2002principles}. 
Since examining the entire population is usually impractical, the researcher selects a subset of the population (a sample) and attempts to estimate a population parameter by statistically analyzing the sample. Probability sampling ostensibly facilitates such statistical generalization \cite[cf.][]{mullinix2015generalizability}. The overwhelming challenge for applying any kind of probability sampling in SE is the absence of comprehensive sampling frames for common units of analysis/observation (see Section \ref{discussion}). This section describes some probability sampling approaches that are relevant to SE research. 

\subsubsection{Whole frame} 
``Whole frame'' simply means that \textit{all} items in the sampling frame are selected. Suppose a researcher wants to assesses morale at a specific software development company. The company provides a complete list of developers and their contact information. The researcher creates a survey with questions about job satisfaction, views of the company, employees' future plans, etc. They send the questionnaire to all of the developers---the entire sampling frame. Whether this is technically ``sampling'' is debatable, but it is an important option to consider, especially when data collection and analysis are largely automated. 

\subsubsection{Simple Random Sampling} 
Simple random sampling means that items are selected \textit{entirely by chance, such that each item has equal chance of inclusion}. Now suppose the above survey finds low morale. The researcher decides to follow up with some in-depth interviews. However, interviewing all 10,000 developers is clearly impractical, so the researcher assigns each developer a number between 1 and 10,000, uses a random number generator to select 20 numbers in the same range, and interviews those 20 developers. This is simple random sampling because the researcher simply chooses \textit{n} random elements from the sampling frame.   

\subsubsection{Systematic Random Sampling} 
In systematic random sampling, given an interval, x, \textit{every x\textsuperscript{th} item is selected}, from a starting point chosen entirely by chance. Suppose, to complement the interviews, the researcher decides to review developers' posts on the company's messaging system (e.g. Slack). Suppose there is no easy way to jump to a random message and there are too many messages to read them all. So the researcher generates a random number between 1 and 100 (say, 47) and then reads message 47, 147, 247, etc. until reaching the end of the messages. This is systematic random sampling. Each post still has an equal probability of inclusion; however, the consistent interval could bias the sample if there is a recurring pattern that coincides with the interval (e.g. taking annual weather data in the middle of summer vs. the middle of winter).

\subsubsection{Panel sampling} 
Panel sampling is when the same sample is studied two or more times. Now suppose the researcher implements a program for improving morale, and a year later, re-interviews the same 20 developers to see if their attitudes have changed. This is called \textit{panel sampling} because the same \textit{panel} of developers is sampled multiple times. Panel sampling is probability sampling if the panel is selected randomly, and non-probability sampling otherwise. 

\subsubsection{A repository mining example of probability sampling} All four of these probability sampling strategies could also be applied in, for example, repository mining. We could (in principle) study every public project on GitHub using GHTorrent \citep{Gousios2013} as our sampling frame (whole frame), randomly select 50 projects (simple random sampling), sort projects by creation date and study every 100th project (systematic random sampling), or take repeated measurements from the same 100 projects over time (panel sampling).

\subsection{Multistage sampling}

Methodologists often present multistage sampling as a special case where two or more sampling strategies are intentionally combined \cite[e.g.][]{valliant2018designing}. Two common approaches are stratified/quota sampling and cluster sampling.

\subsubsection{Stratified/Quota sampling} 
In stratified and quota sampling, the sampling frame is divided into sub-frames with proportional representation. Suppose that the developer morale survey discussed above reveals significant differences between developers who identify as white and those who do not. However, further suppose that 90\% of the developers are white. To get more insight into these differences, the researcher might divide developers into two \textit{strata}---white and non-white---and select 10 developers from each strata. If the developers are selected randomly, this is called \textit{stratified random sampling}. If the developers are selected purposively, it is called \textit{quota sampling}.  This sampling strategy is interesting because it is \textit{intentionally} non-representative \citep{trost1986statistically}. 

We conceptualize these strategies as multistage because the researcher purposively chooses the strata (stage 1) before selecting the people or artifacts to study (stage 2). This paper uses stratified random sampling (see Section~\ref{method}).

\subsubsection{Cluster sampling} 
Cluster sampling is when the sampling frame is divided into groups and items are drawn from a subset of groups. Suppose that the company from our morale survey example has 20 offices spread around the world. Traveling to all 20 offices for face-to-face interviews could be prohibitively expensive. Instead, the researcher selects three offices (stage 1) and then selects 7 participants in each of these offices (stage 2). This is called \textit{cluster sampling}. If and only if \textit{both} selections are random, it is called \textit{cluster random sampling}. Cluster sampling works best when the groups (clusters) are similar to each other but internally diverse on the dimensions of interest.

If the researcher found that the seven developers at one office seem much happier than developers in the rest of the company and therefore conducted extra interviews at that office, in hopes of unraveling the sources of improved morale, this is called \textit{adaptive cluster sampling} \citep{thompson1990adaptive,turk2005review}.

\subsection{Sampling in qualitative research}
Qualitative researchers have to select both sites (e.g. teams, organizations, projects) and data sources (e.g. whom to interview, which documents to read, which events to observe). Different qualitative research traditions (e.g. grounded theory, phenomenology, ethnography) 
talk about this ``selection'' in significantly different ways~\citep{gentles2015sampling}. Some qualitative researchers use the term ``sampling'' \citep[e.g.][]{glaser2017discovery}. Others argue that ``sampling'' implies statistical generalization whereas qualitative research involves \textit{analytical generalization} \citep{Yin:2017tf,van2016phenomenology}. Others argue that there are many kinds of generalization, and qualitative researchers generalize from data to descriptions to concepts to theories, rather than from samples to populations \cite[e.g.][]{lee2003generalizing}. 

This paper tries to clarify that sampling does not imply statistical generalization. 
Predominately qualitative approaches including case studies, interview studies, grounded theory and action research typically use \textit{non-probability sampling} to support \textit{non-statistical generalization} from data to theory \citep[cf.][]{Stol:2016wm,Checkland:1998ii,ralph2019toward}. Predominately quantitative studies, especially questionnaire surveys, \textit{sometimes} use \textit{probability sampling} to support \textit{statistical generalization} from samples to populations. As we shall see below, however, many quantitative studies also adopt non-probability sampling.  

In our view, selecting sites and data sources is a kind of sampling. Selecting a site because it possesses an interesting property is purposive sampling. Interviewing whoever will speak on a subject of interest is convenience sampling. 

As fledgling concepts or theories begin to emerge, however, the researcher may use them to decide what to focus on next. In the grounded theory literature, \textit{theoretical sampling} refers to selecting items to study based on an emerging theory \cite[e.g.][]{breckenridge2009demystifying,draucker2007theoretical}. For example, suppose the researcher from our running example begins generating a theory of developer morale, which includes a preliminary category ``interpersonal conflict.'' Selecting peer code reviews with many back-and-forth interactions, because they might contain evidence of interpersonal conflict, would be theoretical sampling.

\subsection{Sampling Frameworks and Algorithms}
\label{sec:ConeOfSampling}

Historically, sampling strategies were described using a three-tier framework: population, sampling frame, sample. For example: 

\begin{itemize}
    \item Population: Households in City \textit{X}
    \item Sampling Frame: City \textit{X}'s phone book
    \item Sample: 100 phone numbers randomly selected from City X's phone book 
\end{itemize}

In other words, the sample is a subset of the sampling frame and the sampling frame is a subset of the population. The trouble is, many sampling approaches used in SE research do not map cleanly into this framework. 

For instance, suppose that we are interested in non-code documents in software projects (e.g. specifications, lists of contributors, diagrams, budgets). Furthermore, we are especially interested in documents for open source systems and we have had good experiences mining GitHub, so we will limit ourselves to open source projects on GitHub. We only speak English, so we exclude all non-English documents. Now we randomly select 50 English-language open source projects on GitHub and then randomly select up to ten documents from each selected project. We end up with 500 documents. Now suppose we also contact the owners of each selected project to ask if they object to the research, and suppose two of them do, so we delete the corresponding 20 documents. 

The above example has seven sets. Mapping seven sets into a three-tier framework is intrinsically confusing. It's not clear if the sample is the 500 documents we collected or the 480 we retained. It's not clear if the sampling frame is all GitHub projects or GitHub projects that have non-code documents, that are open source, that are in English or some combination thereof. It is not clear if the population is all non-code documents in the world or limited to GitHub, English, open source projects or some combination thereof. 

Furthermore, while simple random sampling was \textit{involved}, it came after eliminating most software projects. Does claiming that a sample is good because we used probability sampling make any sense if we have previously excluded 99\% of the population? 

We suggest bypassing all this confusion by thinking of most quantitative studies as having a multistage sampling algorithm. Precisely defining what is the sampling frame and what is the sample is not as important as describing the sampling procedures clearly enough to enable others to replicate and understand how the sample might be biased. Authors should identify the population they want to generalize to and papers should reflect on how each step in the sampling algorithm could bias the results. 

If the sample is retrieved using a script, the script should be provided in an appropriately archived replication package. Alternatively, if the sampling is performed manually, the precise, step-by-step algorithm used to determine the sample should be provided either in the main text (for shorter algorithms) or the replication package (for longer algorithms). In studies with many steps (e.g. systematic literature reviews) consider a diagram showing how many items were found or eliminated in each step \cite[cf.][Fig 1]{moher2009preferred}. 

For some studies, sampling algorithms can be expressed in a single sentence such as `we selected Corporation Y because of our pre-existing relationship' or 'we recruited a convenience sample of students from the authors' university using our faculty's student email list.' There is nothing wrong with having a simple sampling algorithm as long as it is appropriate for the goals of the study. 

In contrast, the process by which qualitative researchers choose which data sources to focus on can be complex, intuitive and impossible to describe algorithmically. Rather, qualitative researchers should describe their approach and reasoning. 


\subsection{Representativeness} 
\label{represent}

\cite{kruskal1979representativeI} argue, with extensive examples, that the term ``representative'' has been (mis)used in at least five ways: 
\begin{enumerate}
    \item as a ``seal of approval bestowed by the writer''
    \item as the ``absence of selective forces in the sampling''
    \item as a ``miniature or small replica of the population''
    \item as a claim that its members are ``typical of the population'' or ``the ideal case''
    \item as a claim to heterogeneity or that all subpopulations or classes are included (not necessarily proportionately) in the sample 
\end{enumerate}
Recall: \textit{representativeness} is the degree to which a sample's properties (of interest) resemble those of a target population. Representativeness is a mutual property of a sample and a population. Representativeness is \textbf{not} a property of a sampling technique. This section discusses common misunderstandings of representativeness and arguments for representativeness. 

First, \textbf{representativeness is rooted in positivist epistemology}. Postmodernists, interpretivists, and constructivists reject the entire notion of statistical generalization on numerous grounds, including:

\begin{itemize}
\item Broadly applicable (``universal'') theories of social phenomenon simply do not exist \citep{duignan2014postmodernism}.
\item Each context is unique; therefore, findings from one do not translate wholesale into others  \citep{guba1982epistemological}.
\item Statistical generalization precludes deep understanding of a particular social, political economic, cultural, technological context \citep{thomas2015anatomy}.
\end{itemize}

Representativeness of a sample (or site) is therefore not a valid evaluation criterion under these epistemologies. 
Contrastingly, in positivism, falsificationism and Bayesian epistemology, the primary purpose of sampling is to support statistically generalizing findings from a sample to a population. Representativeness is therefore the overwhelming quality criterion for sampling: good samples are representative---bad samples are biased. 

Second, \textbf{representativeness is widely conflated with randomness.} Suppose that two researchers, Kaladin and Shallan, have a sampling frame of 10,000 software projects, with an average size of 750,000 lines of code: 70\% open source and 30\% closed source. Kaladin randomly selects 10 projects. Suppose 4 of them are open source and 6 are closed source, with an average size of 747,823 lines of code.  Meanwhile, Shallan inspects a histogram of project size and discovers a bi-modal distribution, with clear clusters of large and small projects. Shallan purposively selects 7 large open source projects, 7 small open source project, 3 large closed source projects and 3 small closed source projects. 

This example illustrates several key points about representativeness:
\begin{enumerate}
\item Representativeness is dimension-specific. A sample can be representative on one parameter but not another.
\item Probability sampling does not guarantee representativeness on the dimensions of interest.
\item Non-random samples can be representative on the dimensions of interest.
\item Non-probability sampling can lead to more representative samples than probability sampling.
\item The representativeness of a random sample depends on sample size. 
\end{enumerate}

If an unbiased, random sample is large enough, the law of large numbers dictates that its parameters will converge on the parameters of the sampling frame. What constitutes ``large'' depends on many factors including the strength of the effects under investigation, the type of statistical tests being used and the number of variables involved. The parameters of small samples, however, may differ greatly from the sampling frame. While Bayesian approaches handle small samples more systematically and transparently, establishing representativeness with small samples remains problematic \citep{mcelreath2020statistical}. 

Moreover, no sample size can overcome bias in the sampling frame. Suppose that Kaladin and Shallan want to draw inferences about all of the software projects conducted in Brazil. However, suppose the sampling frame is a list of all \textit{public sector} projects in Brazil. Further suppose that public sector projects are more often large and open source than private sector projects, so Kaladin's sample is biased not only toward open source projects but also toward larger projects, and Shallan's sample is less representative than it first appeared.

Clearly then, \textit{random} is not equal to \textit{representative}. \textbf{Rather than defining representativeness, randomness is one of several possible arguments that a sample \textit{should be} representative.} Extending \cite{kruskal1979representativeI}, we see seven main arguments regarding representativeness (see Table \ref{tab_arguments}). None of these approaches guarantee a representative sample, but each has some merit.

\input{tables/arguments}

As discussed above, we can argue that the sample is representative because individuals were selected randomly. However, random selection is only reliable if the sample size is large and the sampling frame is unbiased, and even then only produces a representative sample most of the time.

Alternatively, suppose we are surveying developers about their perceptions of the relationship between agile practices and morale. We can argue that the larger and broader our sample, the less likely it is to have missed an important subpopulation. The breadth argument supports generalization of correlations (e.g. between agility and morale). Breadth is the argument of heterogeneity sampling, and can apply to convenience and snowball sampling where oversampling some subpopulations is a key threat. Using multiple sampling frames makes the breadth argument more convincing \citep[cf.][]{de2015investigating}.

However, the breath argument does not support point estimates. Suppose only 1\% of our sample reports abandoning agile practices because of morale. 
While the point estimate of 1\% is not reliable, abandoning agile practices over morale issues is probably rare. It seems highly unlikely that a survey of 10,000 developers from 100 countries, including thousands of companies and dozens of industries, would miss a large subpopulation of low-morale agile-abandoners.

Another reasonable argument for representativeness is that sample parameters mirror known distributions of population parameters. If we know the approximate distributions for a population of projects' size, age, number of contributors, etc., we can compare the sample parameters (e.g. using the chi-square goodness-of-fit test) to see if they differ significantly. If the sample parameters are close to known population parameters, the sample is representative on those dimensions. If the sample and population match on known dimensions, it seems more likely (but not guaranteed) that they will also match on unknown dimensions. This is the argument of quota sampling. Unfortunately, concerted efforts will be needed to establish population parameters for common units of analysis in software engineering research. 

A quite different argument is the appeal to universality. Suppose we have good reasons to believe that the phenomenon of interest affects the entire population equally. For example, Fitt's law predicts the time required to point at a target based on the target's size and distance \citep{Fitts:1954wz}. Insofar as Fitt's law is universal, researchers can argue that sampling is irrelevant---all samples are representative. 

The appeal to universality is related to the ongoing debate about generalizing from student participants to professionals \cite[cf.][]{Sjoberg:2002je,feldt2018four}. While the details of this debate are beyond the scope of this paper, it is important not to imply a sample of participants is representative simply because they are professionals. Suppose researchers conduct an experiment based on a convenience sample of six American white, male, professional developers, who have bachelor's degrees in software engineering and are between the ages of 30 and 40.
Implying that this sample is representative of professional developers in general because the participants are professionals instead of students is plainly incorrect. 

Contrastingly, some studies address sampling concerns by simply dismissing them on philosophical grounds (as described above). Others argue that, practically speaking, statistical generalization is not the purpose of the present study \cite[cf.][]{stol2018abc}. There is nothing intrinsically wrong with this as long as the manuscript does not imply representativeness elsewhere. 

Many studies, especially laboratory studies, simply ignore sampling concerns. Sometimes a single sentence acknowledges the limitation that the results may not generalize (see also Section~\ref{sec:justifications}).
Other studies give dubious arguments for representativeness \citep{kruskal1979representativeIII}.

%% file: tables/primerSummary.tex

\begin{table*}
\caption{Summary of Sampling Approaches\label{tab_primerSummary}}
\centering
\begin{tabularx}{\linewidth}{llX}
\toprule
Type & Approach & Capsule Description \\
\midrule        
Non-Probability & Convenience & Select items based on expediency \\
Non-Probability & Purposive & Select items most useful for study's objective \\
Non-Probability & Referral-chain & Select items based on relationship to existing items \\
Non-Probability & Respondent-driven & Bias-mitigating variant of referral-chain \\
Probability & Whole frame & Select the entire sampling frame \\
Probability & Simple random & Select items entirely by chance \\
Probability & Systematic random & Select every xth item from a random start \\
Multi-stage & Stratified & Select items from different groups randomly but in equal proportion \\
Multi-stage & Quota & Select items from different groups purposively but in equal proportion \\
Multi-stage & Cluser & Select items in stages, where each stage is a subset of the previous \\
\bottomrule   
\end{tabularx}
\end{table*}

%% file: tables/arguments.tex

\begin{table*}
\caption{Reasonable Arguments regarding Representativeness \label{tab_arguments}}
\centering
\begin{tabularx}{\linewidth}{p{0.11\linewidth}p{0.06\linewidth}XX}
\toprule
Name & GPRS* & Argument & Threats \\
\midrule        
Random \linebreak Selection & No & Each item in sampling frame has an equal probability of inclusion. & Assumes large sample size, good sampling frame and no coincidences. \\
Size & No & Sample is so large that missing a subpopulation is unlikely. & Bias toward one or more subpopulations or on important parameters. \\
Breadth & No & Sample captures a large variance on important parameters. & Point estimates unreliable; groups may not be proportional. \\
Parameter Matching & No & Sample and population parameters have similar distributions. & Possible bias outside of considered parameters; requires known population parameters. \\
Universality & No & All possible samples are representative because the phenomenon affects the entire population equally. &  Phenomenon is not actually universal. \\
Postmodern Critique & No & The entire logic of statistically generalizing from a sample to a population is flawed. & Statistical generalization not supported. \\
Practical \linebreak Critique & No & Generalizing to a population is not the purpose of this kind of study (e.g. case study, experiment). & Statistical generalization not supported. \\
\bottomrule   
\end{tabularx}
\flushleft *Guaranteed to Produce a Representative Sample?
\end{table*}

%% file: tex/3_method.tex
\section{Method} 
\label{method}

To investigate the state of sampling in software engineering, we conduct a critical review. For the purposes of this paper, a \textit{critical review} is similar to a systematic review, except for two key differences: 
\begin{enumerate}
    \item A systematic review typically aggregates evidence regarding causal relationships to generate evidence-based recommendations, whereas a critical review critically evaluates issues.
    \item A systematic review typically aims to collect \textit{all} relevant primary studies on a specific topic \citep{Kitchenham:2007vo} while a critical review analyzes a sample of primary studies sharing some key characteristic(s). 
\end{enumerate}
Critical reviews in software engineering often investigate methodological topics; for example, how grounded theory and ethnography are reported \citep{Stol:2016wm,zhang2019ethnographic} or how qualitative research is synthesized \citep{huang2018synthesizing}. 

To conduct the critical review, we manually retrieved and analyzed a sample of software engineering papers. This section describes the study's research questions, data collection and data analysis. For this study, we adopt a critical realist philosophy.

\subsection{Objective and research questions}
\label{objective-rqs}
The objective of this study is to investigate the sampling techniques used in software engineering research, and their relationship to research methods and units of analysis/observation. This objective motivates the following research questions. In recent, high-quality software engineering research \dots

\begin{description}
    \item{RQ1:} \dots what \emph{sampling approaches} are most common?
    \item{RQ2:} \dots how do authors \emph{justify} their sampling approaches?
    \item{RQ3:} \dots what \emph{empirical research methodologies} are most common?
    \item{RQ4:} \dots what \emph{units of observation} are most common?
\end{description}

RQ1 is the main focus of the study. We include RQ2 to highlight the diversity of arguments for representativeness seen in the literature. Meanwhile RQs 3 and 4 are essential for informing guidance: which sampling approaches are appropriate strongly depends on the research method and unit of observation. Here, \textit{research methodology} denotes the approach to collecting and analyzing data (e.g. controlled experiment, case study, systematic review). 
We decided to concentrate on \emph{units of observation}, rather than units of analysis, because most of the first 20 articles we coded clearly identified their unit(s) of observation, i.e. what was observed/measured/analyzed, whereas identifying units of analysis required substantial interpretation.
For example, if a paper analyzed the commit histories of a sample of open source projects, the units of observation would be the commits themselves. The units of analysis, however, could be the developers contributing to those projects, the projects, both, or something else---depending on how the commits were aggregated and analyzed.

We focus on recent research because research methodology has evolved rapidly in the software engineering community; e.g. increasing use of empirical methods in general \citep{zannier2006success,theisen2018software} and qualitative methods in particular \citep{Stol:2016wm}.  

\subsection{Sampling strategy and inclusion/exclusion criteria}




Answering the above research questions requires manual analysis. Since most SE articles involve sampling, and there are hundreds of thousands of SE articles, we cannot manually analyze the entire population---we need a sample. Since we only report descriptive statistics, we cannot use power analysis to work backwards from estimated effect size to required sample size. Instead, we work backwards from feasibility: based on our research questions and experience, we estimated that it would take about 20 minutes to analyze each paper. Therefore, analyzing 120 articles should take approximately $20*120/60=40$ hours. (We were optimistic; it actually took 15-30 minutes to code each paper with longer discussions of difficult cases.)

Furthermore, we wanted to know where the field is headed. This suggests focusing on recent papers in the most influential outlets. Consequently, we limit our sampling frame to articles published between 2014 and 2019 inclusive in one of the four A* SE outlets according to the 2020 CORE rankings\footnote{\url{https://www.core.edu.au/conference-portal}}:

\begin{enumerate}
    \item The International Conference on Software Engineering (ICSE)
    \item Foundations of Software Engineering (FSE), which was held jointly with European Software Engineering Conference in 2015, 2017, 2018, and 2019
    \item IEEE Transactions on Software Engineering (TSE)
    \item ACM Transactions on Software Engineering and Methodology (TOSEM)
\end{enumerate}

Moreover, we applied several inclusion and exclusion criteria:
\begin{enumerate}
    \item Include: Only full papers.
    \item Exclude: Front matter, letter from editor, etc.
    \item For FSE and ICSE: Include only papers in the main technical track (for symmetry, since TSE and TOSEM do not have an equivalent of workshops, posters, etc.). 
    \item Exclude ICSE and FSE papers for which an extended journal version is already in the sample (no such papers were included so this criterion was never applied).
\end{enumerate}

We did not evaluate the quality of the articles because we care about their sampling technique, not their results.
We wrote a Python script\footnote{\url{https://github.com/sbaltes/dblp-retriever}} to retrieve and validate DBLP\footnote{\url{https://dblp.uni-trier.de/}} metadata of articles published between 2014 and 2019 in either of the four target venues.
The collected information includes venue, year, title, authors, length (pages), session/issue name, and a DOI link to the article PDF.
Using the collected data, the first author manually applied the above-mentioned criteria, resulting in a sampling frame of 1,830 full papers. 

Next, we applied stratified random sampling; that is, we used a true-random number generator\footnote{\url{https://www.random.org/}} to randomly select five papers from each outlet-year (e.g. five papers published by TOSEM in 2016). This means that each outlet and each year have equal representation in our sample. The tool that we implemented to retrieve the papers and the R script implementing our sampling approach are available in our replication package (see Section \ref{sec:DataAvailability}).

Like many SE studies, we adopt a poorly-understood sampling approach. First, we purposively selected ``full research papers published in four good outlets over six years'' and then randomly selected items to study from this more manageable list. We can make the randomness argument to representativeness; however, the four venues in the sampling frame are obviously not representative of all SE research because they are among the most competitive and are all in English. Other researchers might have chosen a different sampling frame. \textit{There is no objective basis on which to select outlets or to study six years of four outlets vs. four years of six outlets vs. one year of 20 outlets.} To proceed, we must simply make reasonable choices and explain their implications.

\subsection{Data extraction and analysis}
Papers were randomly ordered and assigned a unique identifier from 1 to 120. Below, we use \textit{PXX} to denote the \textit{XX}th paper (e.g. P35 is the 35th paper). 

The first author reviewed each paper and recorded, in addition to the automatically collected metadata, the following data points in a spreadsheet: relevant quotations (paper summary, sampling description, quality attributes, sampling limitations, population), number of studies reported in the paper, empirical method, number of samples, sampling stages, sample origin, units of observation, properties analyzed, study population, sampling frame, sampling frame size, sample size, number of items studied. (We created the simple taxonomy of empirical methods shown below inductively.) The complete list of primary studies and dataset are included in the supplementary material (see Section \ref{sec:DataAvailability}). 

Some papers clearly stated the sampling technique, for example:
\begin{enumerate}
    \item ``We invited 739 developers, via e-mail using convenience sampling'' (P35).
    \item ``From the top 500 projects, we sampled 30 projects uniformly at random'' (P43).
    \item ``We used stratified sampling to identify potential interviewees'' (P56).
\end{enumerate}
However, many papers did not clearly explain their sampling technique (specifics below). We therefore had to infer the sampling technique from the text. For example, we inferred purposive sampling from the statement: ``We prepared ... buggy code ... and asked workers to describe how to fix it. Only those who passed this qualifying test could proceed to our debugging tasks'' (P92).

The first author marked ambiguous cases for review by the second author, and these were classified by consensus. For each ambiguous case, we developed a decision rule to guide future classifications (decision rules are included in our replication pack).
We did this in a continuous way, analyzing and resolving ambiguity as we went.

The most important decision rule was to code studies where authors applied certain filtering criteria such as popularity of a project or experience of a developer as \emph{purposive samples}. 
Another important rule was to use the code \emph{sub-sample} to indicate when authors derived a new sample based on a previously introduced one (e.g. selecting developers who were active in projects that had been sampled before).
A third rule was to classify studies based on dominant methodology.
A study that was predominately quantitative with a small qualitative component, for instance, was coded as quantitative. For articles reporting multiple studies where some methods were primarily quantitative and others primarily qualitative, we captured both using a separate row for each study.

%% file: tex/4_results.tex
\section{Results and Discussion} 
\label{results}

We analyzed 120 articles, of which 115 contained an empirical study. Some articles reported multiple studies; some studies had multiple stages of sampling (Table \ref{tab_studied-papers}). We examined the results by \emph{article} (115), \emph{study} (144), and \emph{sampling stage} (210). Note that multiple studies can use the same sample and multiple samples can be used by the same study.

\input{tables/samplingDistribution}

The sample sizes ranged from 1 to 819,000 with a median of 22.
In two cases it was not possible to derive the sample size from the descriptions in the paper.
Only for 60 sampling stages was it possible to derive the size of the corresponding sampling frames.
The sizes of the reported sampling frames ranged from 3 to 2,000,000 with a median of 395.
This section addresses each research question and then comments on observable trends.  

\subsection{RQ1: Sampling techniques used}
Table \ref{tab_strategies} shows the different sampling techniques used in the studies. The frequencies in Table \ref{tab_strategies} total more than 115 because some papers report multiple studies and some studies use multiple samples or sampling stages.

\input{tables/samplingStrategies}

The most common strategies were purposive (149) and convenience sampling (23).
Only 17 stages utilized probability sampling---13 used simple random sampling and 4 used stratified random sampling.
Fourteen stages analyzed their entire sampling frame.
Six stages analyzed generated data, which we exclude in the following.

Eight of the thirteen sampling stages that involved simple random sampling, and three of the four sampling stages utilizing stratified random sampling, were sub-samples of previously derived non-random samples. This is important  because random sampling from a non-random frame undermines the argument that the sample should be representative because it is random (see Section \ref{discussion}) and the coding of papers P16, P27, P35, P43, P47, P48 and P82 in the supplementary material for more details. 

\input{tables/samplingOrigins}

Samples are derived from a variety of sources (see Table \ref{tab_origins}). 
In 62 cases, the sample was based on an existing dataset, typically from related work. In 28 sampling stages, a subset of a larger sample presented in the corresponding paper was used. Thirty-one stages involved sampling from online resources, which was usually done in a purposive manner (27 out of 31 stages). Sources for such samples included the Android developer documentation (P16), online Scala tutorials (P45), LinkedIn groups (P49), app stores (P80, P89, P98), or the online Common Vulnerabilities and Exposures database (P91). Twenty stages involved sampling from the researchers' personal networks (e.g. their students or colleagues) and in 49 cases, the origin of the samples was unclear.
Table \ref{tab_origins} excludes the six papers that used generated data and the five papers that did not report an empirical study.


\subsection{RQ2: Authors' justifications}
\label{sec:justifications}

Of the 120 articles in our sample, 86 provided some justification as to whether their sample exhibits certain quality criteria, often despite a questionable---or unexplained---sampling approach.
The justifications were often mentioned in ``Threats to Validity'' or ``Limitations'' sections.
The most common justification (29 articles) was that the studied artifacts were ``real'' (as in ``real-world'').
P17, for instance, described the sample as containing ``representative real-world ORM applications'', but did not go into details about the actual strategy followed to select those applications.
Further popular adjectives were ``large'' (16), ``representative'' (13), and ``diverse'' (8).

\subsection{RQ3: Research methodologies}

Predominately quantitative studies outnumber predominately qualitative studies 126 to 16. Of those 126 quantitative studies, 98 report experimental evaluations of software tools, models, or machine learning approaches. Nine studies involve mining software repositories; nine studies report on different kinds of user studies, and six are questionnaire-based surveys. 
Beyond that, the diversity of approaches defies organization into common methodological categories. For instance, one study involves comparing software metrics; others build taxonomies.
Table~\ref{tab_cross_method-strategy} contrasts empirical methods with sampling strategy.

\input{tables/crosstab_method-strategy}

\subsection{RQ4: Units of observation}
We organized the primary studies' units of observation into the categories shown in Table \ref{tab_units}. Most of the studies investigate \emph{code artifacts} including GitHub projects (e.g. P4, P17, P43), code commits (e.g. P16, P85), and packages (e.g. P56).
Examples for \emph{other artifacts} include bug reports (P16), faulty rules in model transformations (P29), and test logs (P99).
Besides students (e.g. P69, P94, P96), the category \emph{people} also includes GitHub users (P77), Microsoft developers (P87), and Amazon Mechanical Turk workers (P94).
Table~\ref{tab_cross_method-unit} contrasts units of observation with empirical method.

\input{tables/units}

\input{tables/crosstab_method-unit}

\section{Discussion}

Perhaps the most salient finding of this study is that \textbf{purposive and convenience sampling were the most common sampling strategies} in both qualitative and quantitative studies. Only seventeen articles employed probability sampling and eleven out of the seventeen corresponding sampling stages were sub-samples of non-probability samples.
While non-probability sampling can sometimes produce representative samples, the broad absence of probability sampling suggests that \textbf{software engineering research has a generalizability crisis}.  

While this kind of study cannot determine why probability sampling is so rare, at least three factors may be involved:

\begin{enumerate}
\item Probability sampling is intrinsically difficult for some methodologies (e.g. experiments with human participants).
However, non-probability sampling is popular even for questionnaires and quantitative data analysis studies. 
\item Some SE research adopts interpretivism or other philosophical positions incommensurate with statistical generalization. Although we did not analyze philosophical positions (not least because most articles do not state one), a few studies may fall into this group. 
\item Good sampling frames for most SE phenomena do not exist. 
\end{enumerate}
   
The significance of this third factor cannot be overstated. Without good sampling frames, the randomness argument for representativeness falls apart. We can only claim that the sample represents the sampling frame, but not the population, which is, by definition, what we are trying to generalize to. 

Sampling frames are usually incomplete. Telephone sampling, for example, is incomplete because not everyone has a phone, but \textit{most} households have one or more phones and there are techniques to account for unlisted numbers \citep{LandonJr:1977jf} such as randomly permuting digits of listed numbers or substituting an area code used with mobile phones for an area code used with landlines while retaining the other numbers.  

There is nothing like a phone book of all the software developers, projects, products, companies, test suites, embedded systems, design diagrams, user stories, personas, code faults, or code comments in the world or even in a specific country, language or application domain. Instead, we study samples of GitHub projects (e.g. P10), Microsoft developers (e.g. P87) or Huawei test logs (e.g. P99). If we randomly select enough Microsoft developers, we might get a sample representative of all Microsoft developers, but this is obviously not representative of all developers in the world because even if there were such a thing as an average company, Microsoft would not be it.

The closest we can get to publicly available sampling frames of certain sub-populations of software developers is probably the list of registered Stack Overflow users provided by their official data dump or SOTorrent~\citep{BaltesDumaniOthers2018} and the lists of GitHub projects and users provided by the GHTorrent dataset~\citep{Gousios2013}.
Those datasets, however, both come with their own challenges and limitations~\citep{BaltesDiehl2016}.

This raises two questions: how do we get better samples for our research and how should we assess sampling in an empirical study? We will return to these questions in Section \ref{discussion}. 

Meanwhile, sampling techniques and terminology appear widely misunderstood or misused. The frequency of articles not explaining where their samples came from is concerning. We cannot evaluate sampling bias without knowing the source of the sample. Beyond that, we see the following archetypal problems in the rhetoric around sampling.
\begin{itemize}
    \item Incorrectly using the term ``random'' to mean arbitrary. 
    \item Arguing that a convenience sample of software projects is representative because they are ``real-world'' projects.
    \item Assuming that a small sample is representative because it is random.
    \item Assuming that a large random sample is representative despite being selected from an obviously biased sampling frame.
    \item Implying that results should generalize to large populations without any claim to having a representative sample. 
    \item Dismissing research (particularly case studies) over representativeness concerns when representativeness is not a goal of the study.
    \item Implying that qualitative research is inferior to quantitative research because of the prevalence of non-probability sampling, when most quantitative research uses non-probability sampling.
\end{itemize}

Despite the problems described above, we want to highlight positive examples from recent software engineering papers, not limited to the stratified sample we used for our review.
\begin{itemize}
    \item Some experiments with students present a brief account of how a convenience sample of students was recruited at a certain university, from a certain class or cohort \cite[e.g.][]{mohanani2019requirements}. 
    \item Some case studies mention purposive sampling and state some characteristics of the sites that led to their selection \cite[e.g.][]{ingram2020how}.
    \item Some repository mining studies clearly explain that they randomly draw elements from a purposively chosen sampling frame, acknowledging that their results should statistically generalize to the sampling frame but not the theoretical population \cite[e.g.][]{maalej2013patterns}.
    \item Other repository mining studies clearly explain sampling was constrained by API limitations, and how the researchers worked around those constraints \cite[e.g.][]{humbatova2020taxonomy}.
    \item Some grounded theory studies effectively explain that their goals do not include either representativeness or generalizing from a sample to a population \cite[e.g.][]{sedano2019product}.  
    \item Some questionnaire surveys thoroughly explain their sampling strategy and its implications \cite[e.g.][]{Russo2020Gender}.
    \item Some systematic literature reviews present an extensive step-by-step account of how the primary studies were retrieved and filtered \cite[e.g.][]{BEECHAM2008860}.
\end{itemize}

\subsection{Threats to Validity}
\label{sec:threats}

The preceding results should be interpreted in light of several limitations:

\textbf{Reliability.} Most of the data we extracted from the articles comprises objective, unambiguous facts such as venue, year, title, authors, length (in pages), sample size, and unit of observation. However, some data points (e.g. the methodology or sampling approach) required interpretation and different researchers may have reached different conclusions. We tried to mitigate this threat to reliability by having the second author audit ambiguous cases.

\textbf{Replicability.} To enhance replicability, we provide a comprehensive replication pack (see Section \ref{sec:DataAvailability}) including our data collection scripts, sample and analysis documents. 

\textbf{External validity.} Our results should statistically generalize to our population (recent, high-quality software engineering research) to the extent that the contents of ICSE, FSE, TSE and TOSEM represent high-quality SE research. Generalizability is threatened by the possibility that other venues that publish high quality SE research differ in the sampling approaches that they attract or accept. It is not safe to assume that our results generalize to substantially different venues, research conducted decades ago, or research that will be conducted years in the future. 

\textbf{Construct and measurement validity.} This research does not involve any latent variables, so construct validity does not apply per se. However, the reader could question whether we appropriately ascertain the primary studies' empirical methods, sampling approaches and so on. This kind of study does not lend itself to quantitative analysis of measurement validity. All we can say is that the text of this paper ought to demonstrate that we have sufficient training and knowledge of sampling to be credible.

\textbf{Internal and conclusion validity.} We do not test causal hypotheses so internal validity does not apply. Our analysis is predominately descriptive, so conclusion validity is strong---there are no complicated statistical tests here to go awry. There are no strong assumptions (e.g. homoscedasticity) to be violated.  

%% file: tables/samplingDistribution.tex
\begin{table}
\caption{Properties of studied papers ($n=120$)\label{tab_studied-papers}}
\centering
\begin{tabularx}{\linewidth}{lrrrrrrX}
\toprule
Property & \multicolumn{6}{c}{Count} & Total\\
& 0 & 1 & 2 & 3 & 4 & $\geq$ 5 & \\
\midrule
Studies per paper & 5 & 92 & 17 & 6 & 0 & 0 & studies: 144 \\
Samples per study & 0 & 107 & 26 & 6 & 2 & 1 & samples: 191 \\ 
Sampling stages per sample & 0 & 175 & 26 & 9 & 0 & 0 & sampling stages: 210*\\
\bottomrule   
\end{tabularx}
*includes 6 stages with generated data, which we exclude in the following\\
\end{table}

%% file: tables/samplingStrategies.tex
\begin{table}
\caption{Frequency of sampling techniques ($n=204$ sampling stages)\label{tab_strategies}}
\centering
\begin{tabular}{llr}
\toprule
Type & Strategy & Frequency\\
\midrule
Non-probability	& Purposive	& 149 (73.0\%)\\
Non-probability & Convenience & 23 (11.3\%)\\
Other & Whole sampling frame & 14 (6.9\%)\\
Probability & Simple random & 13 (6.4\%)\\
Probability & Stratified random & 4 (2.0\%)\\
Non-probability	& Snowballing & 1 (0.5\%)\\
\bottomrule   
\end{tabular}
\end{table}


%% file: tables/samplingOrigins.tex
\begin{table}
\caption{Origins of samples in SE research ($n=204$ sampling stages) \label{tab_origins}}
\centering
\begin{tabularx}{\columnwidth}{lXr}
\par
\toprule
Strategy & Definition & Frequency\\
\midrule        
Existing-sample(s) & Study used previously collected data, e.g., reported in related work & 62 (30.4\%)\\
Unclear & The strategy was not explained & 49 (24.0\%)\\
Online-resources(s) & Sample retrieved from the internet (e.g., websites, mailing lists, LinkedIn) & 31 (15.2\%)\\
Sub-sample & Study used a subset of another sample presented in the same paper & 28 (13.8\%)\\
Personal-network & Sample comprises artifacts or people that researchers had access to (e.g., students, industry contacts) & 20 (9.8\%)\\
Other &	For example public or corporate datasets, snowballing. & 14 (6.9\%)\\
\bottomrule   
\end{tabularx}
\end{table}

%% file: tables/crosstab_method-strategy.tex
\begin{table*}
\caption{Empirical method vs. sampling strategy ($n=204$ sampling stages); Strategies: None (sample = sampling frame), Convenience, Purposive, Simple Random.
\label{tab_cross_method-strategy}}
\begin{tabularx}{\linewidth}{Xrrrrr}
\toprule
& None & Con. & Pur. & Sim. & Other \\
\midrule
Experimental tool evaluation (quantitative) & 10 & 8 & 107 & 6 & 2\\
Mining software repositories (quantitative) & 0 & 1 & 8 & 1 & 0\\
Task- \& questionnaire user study (quantitative) & 1 & 3 & 3 & 0 & 0\\
Questionnaire survey (quantitative) & 0 & 2 & 6 & 0 & 1\\
Taxonomy building (qualitative) & 1 & 0 & 3 & 5 & 0\\
Coding of artifacts (qualitative) & 1 & 0 & 4 & 1 & 0\\
Other & 1 & 9 & 18 & 0 & 2\\
\bottomrule   
\end{tabularx}
\end{table*}

%% file: tables/units.tex
\begin{table}
\caption{Units of observation ($n=204$ sampling stages) \label{tab_units}}
\centering
\begin{tabularx}{\linewidth}{p{0.2\linewidth}Xr}
\par
\toprule
Unit & Examples & Frequency\\
\midrule        
Code artifacts & Projects, source code, defects, fixes, commits, code smells & 132 (64.7\%)\\
People & Developers, maintainers, students, interview transcripts & 37 (18.1\%)\\
Non-code artifacts & Bug reports, discussions, pull requests, effort estimates, feature models & 33 (16.2\%)\\
Articles & Papers published in SE journals and conferences & 2 (1.0\%)\\
\bottomrule   
\end{tabularx}
\end{table}


%% file: tables/crosstab_method-unit.tex
\begin{table*}
\caption{Empirical method vs. unit of observation ($n=204$ sampling stages)
\label{tab_cross_method-unit}}
\begin{tabularx}{\linewidth}{X|rrrr}
\toprule
 & Software & Artifacts* & People & Other\\
\midrule
Experimental tool evaluation (quant.) & 101 & 29 & 3 & 0\\
Mining software repositories (quant.) & 8 & 2 & 0 & 0\\
Task- \& questionnaire user study (quant.) & 1 & 0 & 6 & 0\\
Questionnaire survey (quant.) & 3 & 0 & 6 & 0\\
Taxonomy building (qual.)& 5 & 4 & 0 & 0\\
Coding of artifacts (qual.) & 5 & 1 & 0 & 0\\
Other (e.g., mapping studies, Wizard-of-Oz evaluations, think-aloud studies) & 9 & 1 & 18 & 2\\
\bottomrule   
\end{tabularx}
*non-code artifacts
\end{table*}

%% file: tex/5_recommendations.tex
\section{Recommendations} 
\label{discussion}

Section \ref{results} demonstrates that software engineering has a generalizability crisis brought on chiefly by a lack of attention to representative sampling. The following section presents numerous recommendations and guidelines for addressing this crisis.


 


\subsection{Guidelines for researchers}

To improve the conduct and reporting of sampling for any empirical study, we recommend: 
\begin{enumerate}
    \item \textbf{Clarify your philosophical position.} A treatise on 21\textsuperscript{st} century epistemology is not necessary, but one sentence on the study's perspective---positivism, falsificationism, interpretivism, constructivism, critical realism, etc.---will help. Otherwise the reader (or reviewer) may apply inappropriate criteria. 
    \item \textbf{Explain the purpose of sampling.} Clearly state whether you are aiming for a representative sample, or have a different goal (e.g. theoretical saturation).
    \item \textbf{Explain how your sample was selected.} For qualitative studies, the reader should be able to \textit{recover} your reasoning about what to study \citep{Checkland:1998ii}. For quantitative studies, the reader should be able to replicate your sampling approach and the size of the sample should be evident. 
    \item \textbf{Make sure your sampling strategy matches your goal, epistemology, and type of study.} For example, a positivist questionnaire might use respondent-driven sampling; a pilot laboratory experiment might use convenience sampling, and an interpretivist case study might employ purposive sampling. 
    \item \textbf{Avoid defensiveness.} Very few software engineering studies have a strong claim to representative sampling. Overselling the representativeness of your sample is unnecessary and unscientific. For example: 
        \begin{itemize}
            \item Do not misrepresent ad hoc sampling as random.
            \item Do not pretend small samples are automatically representative because they are random or because they were purposefully selected. 
            \item Do not pretend random sample are representative when selected from potentially or obviously biased sampling frames
            \item Do not pretend a sample is representative because it is ``real'' (e.g. professionals instead of students, commercial projects instead of toy examples). 
            \item Do not admit to sampling bias in your limitations section only to pretend the results are near-universal in your conclusion.
        \end{itemize}
\end{enumerate}

Moreover, if representativeness is the goal of the study, we further recommend:
\begin{enumerate}
    \item \textbf{Identify the population;} that is, in principle, who or what you would like to generalize to (e.g. professional software developers in Turkey, code faults in cyber-physical systems).
    \item \textbf{Present your sampling algorithm} (see Section~\ref{sec:ConeOfSampling}). If the algorithm maps neatly into the tripartite model of a population, a sampling frame and a sample, feel free to use those terms. However, if the algorithm has many stages and does not map neatly into the tripartite model, just focus on providing a replicable, concise, algorithmic account of how another person can generate the same sample. It is not necessary to name every stage or identify one specific stage as the ``sampling frame.'' 
    \item If you use data collection scripts, provide them. 
    \item If the sampling strategy has many stages, consider a diagram or flow chart. 
    \item If your sampling approach was nondeterministic (e.g. theoretical sampling), describe your reasoning.  
    \item Give an \textbf{explicit argument for representativeness} (cf. Table~\ref{tab_arguments}). Admit the generalizability threats implied by this argument. 
    \item \textbf{Clearly explain how the sample could be biased.} For complicated sampling strategies, discuss bias for each step in your algorithm or diagram explaining the sampling approach. This could be presented in the sampling section or with limitations.
    \item \textbf{Publish your sample} as part of a replication package \textbf{if and only if} it does not contain sensitive or protected information. Be very careful of the potential for re-identifying de-identified data. 
\end{enumerate}

For example, if we are conducting a randomized controlled experiment with a convenience sample of software developers, we should not claim that the sample is representative because the participants are professionals. We should explain that we are prioritizing internal validity and that this kind of research does not support statistical generalization. Similarly, for a case study, we should explain that the purpose of a case study is to understand one site deeply, not to generalize from a sample to a population. We should not pretend that one or a few sites are representative because they are large companies or real projects or use popular practices. For interpretivist research, meanwhile, we should discuss transferability rather than external validity. 

The situation is trickier for questionnaire surveys, qualitative surveys, repository mining and exploratory data science, where external validity is more often a key criterion. Authors of these studies should offer an explicit argument for representativeness and clearly explain attempts to mitigate sampling bias. 

\subsection{Mitigating sampling bias in different kinds of studies} 

Several other techniques can help improve sampling in certain situations. For large samples, we can use bootstrapping to assess stability. For instance, if we have a convenience sample of 10,000 Java classes, we can randomly exclude 1,000 classes and check whether it perturbs the results. 

If a sampling frame is biased, consider replicating the study using different sampling frames. For example, if we find a pattern in a sample of GitHub projects, replicate the study on a sample of SourceForge projects. The more diverse the repositories, the more likely the results generalize. This could be a full replication or a limited sanity check with a small sample from a different domain. For studies of software projects, consider using sample coverage; that is ``the percentage of projects in a population that are similar to the given sample'' \citep{nagappan2013diversity}, to support heterogeneity sampling.

Developers (or developers with certain experiences) can be treated as a hidden population. If respondent-driven sampling (Section \ref{rds}) can reach injecting-drug users and sex workers \citep{Malekinejad:2008gm}, surely it can help reach software developers. If representativeness is not a priority, \cite{salleh2018recruitment} provide several suggestions for recruiting industry participants including exploiting local practitioner communities, using managers and moderators as recruiters, and using snowball sampling.

Meanwhile, many practices can reduce sampling bias and response bias in \textbf{questionnaire surveys} \citep{dillman2014internet}. These include starting with important questions that resonate with participants (not demographics), avoiding matrix questions, avoiding mandatory questions, and sending reminders. Offering incentives (e.g. cash, prizes) is also effective. Some research suggests that offering charitable donations does not increase response rates \citep{toepoel2012effects}; however, none of this research was done with software developers, and donating to an open source project might be more effective than small cash incentives for motivating open source contributors to complete a questionnaire \cite[e.g.][]{ralph2020pandemic}. There are also myriad techniques for \textit{assessing} response bias \citep{sax2003assessing}. Many questionnaire studies in SE use none of these techniques.

Similarly, sampling bias and publication bias in \textbf{systematic reviews} can be addressed by (i) forward and backward snowballing on references, (ii) searching multiple databases, (iii) searching pre-print servers and dissertations, and (iv) requesting unpublished work in an area through relevant mailing lists, and (v) checking websites of prolific researchers in the area.

Evaluating sampling bias in \textbf{software repository mining} is fraught. 
Each repository is likely biased in unpredictable, non-obvious ways.
Therefore, we cannot safely assume that random samples from one repository are representative of other repositories or software in general.
Comparing samples from multiple repositories may help improve representativeness, or at least assess stability, but open source projects may differ from closed-source projects in unknown ways and the private code that companies are willing to share might systematically differ from the code they will not share \citep{Paulson:2004fl}. Research comparing projects stored in public and private repositories is especially needed, but intrinsically difficult. One (ethically questionable) option is to examine private code that has been leaked. 

In the long term, SE research needs better sampling frames. One way to achieve this is to develop curated corpora like the Qualitas corpus, ``a large curated collection of open source Java systems'' \citep{Tempero:2010fh}. Similar corpora could be developed for many kinds of code and non-code artifacts used in software projects, including design specifications, requirements specifications, diverse models and diagrams, user stories, scenarios, personas, test cases, closed-source Java systems, systems in all the other common languages, unit tests, and end-user documentation. Creating any one of these corpora is a major undertaking and should be recognized as a significant research contribution in itself. Even without good demographic information, the representativeness of a curated corpus can be improved in numerous ways:
\begin{enumerate}
\item Including artifacts from diverse domains (e.g. aerospace, finance, personal computing, robotics).  
\item Including artifacts from diverse software (e.g. embedded systems, enterprise systems, console video games).
\item Making the corpus large enough to support heterogeneity sampling and bootstrapping.
\item Attempting to match the parameters we can discern; for example, we could attempt to include artifacts from different countries according to the size of each country's software industry.
\end{enumerate}
Corpora improve reproducibility because one corpus can support many studies. Furthermore, building corpora helps to separate the difficult task of creating and validating a good sampling frame from any particular study of the items in the corpora. This makes research more manageable. That said, it may not be possible to create a perfect, unbiased corpora. The idea is to create corpora that are \textit{less biased} than other sampling frames, or at least corpora with well-understood biases. Corpora can quickly go out of date and there exists an innate tension between keeping corpora stable and up-to-date. 




\subsection{Guidelines for reviewers}

From our experience, many reviewers struggle to evaluate sampling. Our advice is to \textbf{evaluate sampling in the context of a study's philosophy, methodology, goals and practical realities}. Assessments of external validity or transferability in principle and out of context are unhelpful. The same goes for assessing sampling outside of peer review (e.g. to include primary studies in a systematic literature review). 

For interpretivist studies, it is sufficient for researchers to justify site selection and explain their data collection. Complaining about low external validity in a case study is typically unreasonable because that is not what a case study is for \citep{stol2018abc}.

When reviewing a positivist study that does not aim for generalization (e.g. a laboratory experiment with human participants) only worry about high-level external validity threats such as using student participants instead of professionals. Complaining about low external validity in a laboratory experiment is typically unreasonable because that is not what a lab study is for \citep{stol2018abc}. 

However, when reviewing a study that does aim for generalization (e.g. a questionnaire study) insist on reasonable attempts to mitigate sampling bias. The whole point of a large questionnaire survey is to sacrifice internal validity for external validity. If external validity is the main priority of a study, it should have a defensible claim that its sample is representative. 

For example, suppose we are evaluating a questionnaire study of 3D animators at AAA game companies. The authors recruited animators by posting on mailing lists, which is basically convenience sampling---end of sampling discussion. This should be rejected not because it uses convenience sampling but because appropriate, practical steps for mitigating sampling bias were not taken. Authors should have used respondent-driven sampling, or found a list of AAA game companies and used stratified random sampling, or advertised on multiple channels and compared them \citep[cf.][]{BaltesDiehl2016}. They should have reported response rates or bounce rates, compared early responders to late responders and so on.  

In contrast, suppose we are evaluating a constructivist grounded theory study of agile practices at a Norwegian software company. We could say ``this study has low external validity because we cannot generalize from an \textit{n} of 1.'' This is simultaneously true and inappropriate. External validity is not an appropriate quality criterion for this kind of study \citep{charmaz2014constructing} and statistical generalizing is not its aim \citep{stol2018abc}. Instead, we should be asking why this site was selected, how the researchers went about theoretical sampling, and to what extent the resulting theory seems transferable to other contexts \citep{Stol:2016wm}. 

Assessing sampling in software repository mining is difficult because it comes from SE, so we are creating the norms. What we can say with confidence is randomness argument to representativeness is unconvincing when the sample was randomly selected from a purposively assembled subset of the repository, and there is no evidence that the software in the repository does not differ from software in general on the dimensions of interest.

The key is to evaluate studies against the norms for that particular kind of study, and to question whether researchers applied practical mitigations available to them (like sending reminders for a questionnaire), not to evaluate abstract quality criteria like external validity. 
Pilot and proof of concept studies investigate something under ideal---not representative---conditions. 
For experiments with human participants, representative sampling is often prohibitively expensive. 
Most qualitative research does not seek to generalize to other contexts, so representative sampling is irrelevant, and disparaging ``a sample of 1'' is merely prejudice against qualitative research. 
For studies that do not aim for representativeness, reviewers should instead focus on over-generalization. Lab studies and pilot studies under ideal conditions do not show that something works in real life; qualitative field studies do not establish universality. 

Reviewers should check whether the sampling strategy is commensurate with the study's implications. Non-representative sampling should be accompanied by acknowledging that external validity is limited. Such acknowledgments should not be followed by a sneaky implication that the results are universal. Misusing the term ``random'' should not be tolerated. 

Finally, reviewers should consider whether the sample is large enough. For studies using frequentist statistics to test causal hypotheses, power analysis should be used to estimate desired sample sizes \citep{cohen1988statistical}. Reviewers should also consider local norms, for example average sample sizes \citep{Caine:2016:Local}.

For case studies, there is a longstanding debate in the case study methodology literature about whether a typical manuscript should report one case or multiple cases. Our position is that: (1) reporting multiple, related cases in a single paper is advantageous \citep{Yin:2017tf}; however, single-case studies can still have value and should not be rejected out of hand \citep{easton2010one}. Case studies---single or multiple---do not aim for statistical generalization, and adding a second, third or fourth case will not change that. 

The discussion of representativeness above foreshadows the difficulty of assessing a sampling strategy. The representativeness of a sample is often subjective, and representativeness is not always the goal of the sampling strategy. We suggest the following questions for guiding assessment of a sampling strategy: 

\begin{enumerate}
\item Has the paper specified a philosophical stance?
\item Has the paper specified the goal of the sampling strategy (e.g. representativeness, convenience)?
\item Has the paper described the sample and sampling strategy sufficiently?
\item Is the sampling strategy consistent with the stated goal and philosophical position? 
\item Is the sampling strategy reasonable given the context, constraints and maturity of the research? 
\item Are the limitations of the sample acknowledged?
\item Does the sampling strategy match the paper's knowledge claims?
\item If representativeness is the goal, what argument to representativeness is made? Is it reasonable given the type of study and practical constraints? Are there \textit{specific steps} the researchers could have taken to mitigate sampling bias but did not?
\end{enumerate}

It is very important for a sampling strategy to support the specific knowledge claims of the paper. When an article makes claims about a population, based on sample, or makes generic claims of the form X causes Y, we can and should question the article's argument for representativeness. Much qualitative research, in contrast, seeks to understand one specific context with no attempt to generalize knowledge to other contexts. In such research, challenges to representativeness are far less important. 

This is not to say that a paper should be rejected out of hand because some details are missing or the conclusion overreaches. Reviewers can often simply request clarifications or rewording. Even when representativeness is the goal and reasonable attempts to mitigate sampling bias have not been made, such attempts may be possible in a multi-phase review process.   

The key phrases here are ``reasonable'' and ``practical constraints.'' Any criticism of a paper's sampling approach should include suggesting specific, practical techniques to mitigate sampling bias. Complaining that a systematic review should have addressed sampling bias through reference snowballing is reasonable; complaining that a study of unit tests should have used probability sampling when no reasonable sampling frame exists is not.



\subsection{A Strategy for Addressing the Generalizeability Crisis}

Based on the preceding discussion, we suggest a multi-pronged strategy for improving the generalizeability of software engineering research. 

\begin{enumerate}
    \item \textbf{Education}. PhD students need to take formal research methods courses. There is simply no excuse for failing to teach PhD students a variety of sampling strategies and the technical meaning of ``random.'' 
    \item \textbf{Larger samples.} Many studies are simply too small for their intended purposes. It is not normally possible to detect modest effect sizes with a twelve-participant experiment or to reach theoretical saturation with six interviewees. 
    \item \textbf{More probability sampling.} Probability sampling is not always appropriate, but should be used more often where it is appropriate. Many repository mining studies, for example, would benefit from probability sampling.
    \item \textbf{More sophisticated sampling.} Where possible, researchers should try to replace basic sampling approaches (e.g. snowball) with more sophisticated approaches (e.g. respondent-driven) to mitigate sampling bias. 
    \item \textbf{Multiple sampling frames or approaches.} Drawing samples from multiple sources (e.g. countries, organizations, projects, repositories) not only increases sample heterogeneity but also provides a way to investigate whether results are source-specific. 
    \item \textbf{Develop better sampling frames.} A comprehensive research program is needed to develop better sampling frames and better techniques for accessing elusive populations of common units of analysis. 
\end{enumerate}

Meanwhile, the flip-side of the generalizeability crisis is an over-generalization crisis. Anecdotally, many papers seem to overstate the representativeness of their samples and the generalizability of their findings. An important avenue for future work is to understand how and why SE researcher's overstate generalizability, and what can be done about it.

%% file: tex/7_relatedWork.tex
\section{Related Work} 
\label{sec:RelatedWork}

Many papers and books have been published that discuss sampling. Most general methods textbooks, such a Trochim's excellent \textit{Research Methods Knowledge Base} (\citeyear{trochim2001research}) and SE-specific methods texts \citep[e.g.][]{wohlin2012experimentation,Foster:2014tx,easterbrook2008selecting} discuss the logic of sampling, and many of the strategies described in Section \ref{background}. Some previous papers tackle sampling from the perspective of a particular methodological context such as software repository mining \citep{nagappan2013diversity,cosentino2016findings} or questionnaire surveys \citep{kitchenham2002principles,de2016surveys} while others provide more general primers similar to that in Section \ref{background} \citep[e.g.][]{de2015investigating,kitchenham2008personal}. We extend these works by \textbf{untangling representativeness from randomness} and elucidating the many arguments for representativeness that can be deployed in SE research.

Meanwhile, a few studies have investigated the prevalence of probability sampling in SE research. \cite{amir2018there} found probability sampling in only 13 of 236 articles in the proceedings of the \textit{International Symposium Empirical Software Engineering and Measurement} from 2012 to 2016 inclusive. Similarly, \cite{cosentino2016findings} found that purposive and convenience sampling were the most commonly employed strategies in studies of GitHub projects; while \cite{de2015characterizing} found that purposive and convenience sampling were he most common strategies in questionnaire surveys. This paper confirms the same trend in a method-agnostic sample of papers published at top venues. 

Some existing work analyzes specific issues related to sampling and gives recommendations. For instance, \cite{BaltesDiehl2016} discuss barriers to random sampling for questionnaire surveys, highlighting in particular ethical concerns with ``contacting developers on GitHub using email addresses users did not provide for this purpose'' and the need to establish population parameters to assess representativeness. Many papers discuss the challenges of generalizing from students to professionals \citep[e.g.][]{feldt2018four,falessi2018empirical} and from open-source to closed-source projects \citep[e.g.][]{cosentino2016findings}. 

Other existing work gives various recommendations related to sampling. \cite{stol2018abc} introduces the concept of a ``sample study,'' that is, a study that aims to statistically generalize from a sample to a population; \citeauthor{stol2018abc}'s point, which we echo, is that criticizing the external validity of a study that \textit{does not aim} to support statistical generalizability is nonsensical. \cite{nagappan2013diversity} suggest ``sample coverage''; that is ``the percentage of projects in a population that are similar to the given sample'' as a criterion for evaluating sampling, which is closely related to what we call the breadth argument to representativeness. Similarly, \cite{torchiano2017lessons} suggest assessing representativeness by comparing a sample to what is known about the target population, which is closely related to what we call the parameter-matching argument to representativeness. However, \cite{torchiano2017lessons} share our concern that population parameters may be unknown. 

In contrast, \cite{de2015characterizing} give recommendations for generating and describing sampling frames which are very different from ours. Their recommendations, including recruiting individuals from ``an active and thematic SE discussion group'' and studying ``multiple'' open source projects, do not address the core issues of representativeness and biased sampling frames tackled by this paper. Similarly, \cite{de2015investigating} and \cite{de2016surveys} present a framework for sampling professionals for online surveys and introduce \textit{LinkedIn} as an exemplary ``source of sampling''. This again is very different from our perspective, since our guidelines are not limited to questionnaire-based online surveys and we further agree with \cite{BaltesDiehl2016} that contacting developers using contact information not provided for this purpose is ethically fraught. Nevertheless, there seems to be agreement that better sampling frames are needed and that the sampling procedure should be presented as transparent as possible.

This paper extends previous recommendations by elucidating a six-pronged strategy for addressing the generalizability crisis in SE research. We also suggest \textbf{various approaches for mitigating sampling bias} including treating developers as a hidden population, bootstrapping, selecting projects from multiple repositories (not just GitHub) and charitable incentives for taking questionnaires. This paper also develops specific guidelines for reviewers---a crucial step, as many papers are unreasonably criticized for non-probability sampling when the study does not aim to support statistical generalization, there is no appropriate sampling frame to support probability sampling, and very few of the other studies accepted by the same venue use probability sampling effectively.    

%% file: tex/6_conclusion.tex
\section{Conclusion} 
\label{conclusion}

This paper makes several contributions:

\begin{enumerate}
    \item An introduction to sampling with examples from SE research.  This exposition is more grounded in SE research than previous discussions in reference disciplines, and more general than previous discussions within SE.
    \item An analysis of the state of sampling methods in a stratified random sample of recent SE research in leading venues. It shows that probability sampling is rare, and most probability samples are drawn from unknown or non-representative sampling frames.
    \item A novel exploration of the arguments for representativeness (see Table \ref{tab_arguments}), clarifying that randomness neither equals nor guarantees representativeness, and that a nonprobability sample can be more representative than a probability sample. 
    \item Guidelines for conducting, reporting and reviewing sampling.
\end{enumerate}

A sample is representative of a population insofar as the relevant parameters correspond. We can assume that a random sample is representative if and only if it is sufficiently large and is drawn from an unbiased sampling frame. Few SE studies use random sampling. Of those that do, some samples are too small to assume representativeness and others are drawn from plainly biased sampling frames. Researchers make various arguments for why their samples should be representative: the sample is large, includes diverse items, matches known population parameters, etc.

This creates a paradox: the lack of representative sampling is undermining SE research but rejecting a study over its non-representative sample is capricious because virtually none of the other studies have representative samples. We can escape this paradox by working towards more representative sampling in studies where generalizability is desired. For questionnaires especially, researchers should apply known techniques for mitigating and estimating sampling bias. We also need to develop more curated corpora of SE artifacts, better sampling frames for SE professionals and techniques for mitigating sampling bias in repository mining.

Furthermore, we need reciprocal willingness of researchers to present their research more honestly and reviewers to stop capriciously rejecting work over unavoidable sampling bias. This means no more mislabeling ad hoc sampling as random, no more pretending small samples are automatically representative because they are random, and no more ignoring the potential differences between sampling frames and populations.
It also means no more accepting convenience sampling for experiments while criticizing convenience sampling for case studies and interviews. No more encouraging snowball sampling for literature reviews while rejecting it in questionnaires.

The contributions above should be considered in light of several limitations. We operationalized ``recent high-quality software engineering research'' as articles published in four top venues over six years. Our sample is therefore unlikely to represent the broader field. Studies that were published twice (e.g. a paper in FSE followed by an extended version in TOSEM) have a greater chance of being selected. 
Moreover, the analysis was hindered by widespread confusion regarding sampling techniques and research methodologies.

Additionally, some of the guidelines suggested in this paper are not directly supported by empirical evidence. The guidelines are meta-science, and like most meta-science, are somewhat polemical \cite{ralph2020empirical}. It simply is not practical to conduct experiments to determine whether aligning a study's sampling strategy with its goals, epistemology and methodology improves scientific outcomes. 


In conclusion, we hope that this article's sampling primer, empirical results and recommendations raise awareness of and provide at least some basis for improving sampling in SE research. 


\section{Data Availability}
\label{sec:DataAvailability}
Supplementary materials, which have been archived on Zenodo \citep{baltes2020SamplingSupplement}, include:
\begin{itemize}
    \item An Excel spreadsheet containing the complete list of articles, all of the extracted data and all of our analyses; 
    \item The scripts we used to retrieve sampling frame and sample.\footnote{a more recent version of \textsf{dblp-retriever} may be available at \url{https://github.com/sbaltes/dblp-retriever}}
\end{itemize}